\documentstyle[12pt]{article}
\newcommand{\barbox}{\stackrel{-}{\Box}}

\begin{document}
\baselineskip=0.7truecm
\begin{titlepage}
\title{MAXIMAL ACCELERATION OR MAXIMAL ACCELERATIONS?}
\author{Antonio Feoli\thanks{E-mail: feoli@unisannio.it}
 \\Dipartimento d'Ingegneria, Universit\'a del Sannio,\\
Corso Garibaldi n. 107, Palazzo Bosco Lucarelli\\ 82100 - Benevento -
Italy}
\date{\empty}
\maketitle
\begin{abstract}
We review the arguments supporting the existence of a maximal
acceleration for a massive particle and show that different values of
this upper limit can be predicted in different physical situations.
\end{abstract}

\end{titlepage}

\section{Introduction}

Since Weyl's attempt [1] to unify Gravitation and Electromagnetism,
many different generalizations of Einstein Relativity have been
proposed increasing the number of dimensions as in Kaluza - Klein
approach or using Finsler spaces, complex manifolds,  scalar - tensor
coupling, extended particles in the form of strings or bubbles, etc.
Some of these models start from fixing another observer independent
scale in addition to the
 speed of light of standard Special Relativity. Recently a new
interest in this kind of theories has been revived by Amelino-Camelia
[2] who proposed a model based on the existence of a Length scale and
by Magueijo and Smolin [3] who preferred to use an Energy scale. Both
the attempts lead to a resulting maximal momentum (which has been
found also by Low [4] using a different approach). Also Ahluwalia and
Kirchbach [5] argued that the interplay of gravitational and quantum
realms requires two invariant scales and obtained a gravitationally
modified de Broglie wavelength that acquires a value of the order of
Planck length in  Planck regime. Another way was followed by Ketsaris
[6] who, starting from a seven dimensional manifold, obtained a
Maximal Acceleration and a Maximal Angular Velocity. The three models
of "Quantum Special Relativity" need to be developed and
Amelino-Camelia [7] suggests "to reach a formulation of a {\it
Quantum General Relativity} by an appropriate extension of the k -
Minkowski spacetime to some sort of k - phase space (which however
here is intended as the space $ x_i, t, p_i, E$ rather than just  $
x_i, p_i$)".  This was the same aim of the "Quantum Geometry"
proposed by Caianiello [8] more than twenty years ago that led to the
introduction of a maximal acceleration [9]. The boosts deformations
of a "Quantum Special relativity" with an upper limit on the
acceleration were derived by Scarpetta [10] in 1984 and even
different models of "Quantum General Relativity"  have already been
developed by Brandt [11] and Schuller [12]. Caianiello himself and
his co-workers have analyzed the quantum corrections to the classical
spacetime metrics due to the existence of a Maximal Acceleration. His
idea [13] is  that the simplest theoretical framework, which includes
maximal proper acceleration,
 consists of considering as physical
invariant, not the classical four dimensional space time distance
element, but a new one, more general, defined in an eight dimensional
phase space, where the infinitesimal element of distance can be
written $$ d\tilde{s}^2=g_{AB}dx^Adx^B=g_{\mu\nu}dx^{\mu}dx^\nu
+{c^4\over A^2_{max}} g_{\mu\nu}d\dot x^{\mu}d\dot x^\nu. \eqno(1)$$
where $\dot x^{\mu}={dx^{\mu} / ds}$ is the relativistic
four--velocity. The consequence is that a particle of mass $m$
accelerating along its worldline, behaves dynamically as if it were
embedded in a spacetime of metric $$ d\tilde{s}^2 = ds^2 \left( 1 -
{{ c^4 |g_{\mu \nu} \ddot{x}^{\mu} \ddot{x}^{\nu}|}\over{A_{m}^{2}}}
\right)\,, \eqno(2)$$ But which is the right value of $A_{m}$ to use
in (2)?  From the historical point of view the  maximal proper
acceleration has been first derived starting from the principles of
Quantum Mechanics and Relativity by Caianiello [14] who obtained the
value $$A_{max} = {2 m c^3 \over \hbar} \eqno(3)$$ {\it depending on
the rest mass of the particle $m.$} While in the context of quantum
geometry [8] the maximal acceleration is generally referred to
extended particles, the proof [14], from the Heisenberg uncertainty
principle, holds also for point particles. Starting from the value
(3), we obtained interesting results both with a simplified model
(lacking of covariance) applied to Rindler [13], Schwarzschild [15],
Reissner - Nordstrom [16], Kerr [17] and Robertson - Walker [18]
metrics, and with a fully covariant approach that leads to a complete
integrability of equations of motion (up to now) only in spacetimes
of constant curvature [19].
 As the concept of maximal acceleration  has proved to be very
fertile producing a lot of different interesting models, in this
paper we want to review critically the main arguments that support
the {\it existence} of this upper limit but not its {\it uniqueness}.
We will quote old results and will give a new interpretation to some
recent papers showing that, after Caianiello's proposal, different
values of Maximal Acceleration can be  predicted in different
physical situations.

\section{Maximal acceleration for extended objects}

It is well known that massive extended objects imply critical
accelerations, determined by the extension of the particles and by
the causal structure of the space--time manifold. For instance, in
classical relativity [20], an object of proper length $\lambda$, in
which one extreme point is moving with acceleration $a$ with respect
to the other, will develop a Rindler horizon at a proper distance
$a^{-1}$ from the accelerated  extremity, so that all parts of the
object can be causally connected only if $\lambda  <  a^{-1}$. This
implies a proper critical acceleration $a_c \simeq \lambda^{-1}$
which depends on $\lambda$ and diverges in the limit in which the
object reduces itself to a point--like particle.

In the quantum relativistic context, the analysis of string
propagation in cosmological backgrounds revealed that an acceleration
higher than the critical one give rise to the onset of Jeans--like
instabilities [21] in which the string oscillating modes develop
imaginary frequencies and the string's proper length diverges.
Gasperini [22] has given a very interesting kinematic interpretation
of this string instability, showing that it occurs when the
acceleration induced by the background gravitational field is large
enough to render the two string extremities causally disconnected,
because of the Rindler horizon associated with their relative
acceleration. This critical acceleration $a_c$ is determined by the
string  size $\lambda$ and is given by $a_c = \lambda^{-1} = (m
\alpha^{\prime})^{-1}$ where m is the string mass and $\alpha^{\prime
-1}$ the usual string tension.

 Frolov and Sanchez [23]  analyzed the dynamics of an  uniformly accelerated open string
in flat space. They used the classical Rindler metric $$ ds^{2}  =
-\xi^{2}    d \eta^{2}    +   d   \xi^{2} +   d y^2   + d
z^{2}\eqno(4)$$ (where the Rindler coordinates are $   \xi = 1/a,~~
\eta     =   a   s $, and   $  a   $   is  the acceleration) and
supposed that there are two heavy particles (e.g. monopoles) at the
ends of the string, numbered with the indices $1$ and $2$, on which
some external force is applied in such a way that both particles are
moving with the same constant proper acceleration $a=g$. In an
inertial frame of reference the coordinates $(t, x, y, z)$ are chosen
in such a way that the $x$--axis coincides with the direction of
acceleration, while the $y$--axis is parallel to the distance L
between the ends of the string; correspondingly, the Rindler
coordinates are $(\eta, \xi, y, z)$ and in the accelerated Rindler
frame the particles at the string ends obey the boundary conditions:
$$   \xi_{1}   =   \xi_{2} = g^{-1};   \;  \; \;
    y_{1}   = -y_2  =   L/2;  \;    \;  \
    z_{1}   =   z_{2}   =0  \eqno(5)$$

  Putting $y  =   L\sigma/\pi$, the spatial parameter $\sigma$
varies from     $-\pi/2$   to $\pi/2$.

They found
 a  special solution of the equations of motion
 describing an uniformly accelerated     string, which moves as a
 rigid  body    without any excitation:
$$\xi   =   {L\over \pi \beta} \cosh \left({\beta\pi\over L}y\right)
\eqno(6)$$   As for the $\beta$ parameter, its value is fixed by the
boundary condition that the string ends must move with the assigned
acceleration $g$, expressed by the equation: $$\cosh
\left({\beta\pi\over 2}\right) = {\beta\pi\over gL} \eqno(7)$$ For
different values of the acceleration $g$, this equation admits two,
one, or no solution for $\beta$. Frolov and Sanchez  proved that
rigid equilibrium configurations of the accelerated string exist only
for an acceleration less than a critical one. Finally, they
calculated the string size $\lambda$ in the Rindler frame: $$ \lambda
= {2L\over\beta\pi}\sinh\left({\beta\pi\over 2}\right)=
{2L\over\beta\pi}\sqrt{\cosh^2 \left({\beta\pi\over 2}\right) - 1}
\eqno(8) $$

 Now, in order to compare their result with Gasperini's one,  we can substitute
  (7) in (8) and easily obtain the
parameter $\beta$ in terms of the acceleration and the proper length
of the string: $$ \beta = \pm \frac{2 gL}{\pi \sqrt{4 - g^2
\lambda^2}} \eqno(9)$$ From the equation above we calculate that
$\beta$ is real if $$g < \frac{2}{\lambda} \eqno(10)$$ that confirms
the critical value of acceleration predicted by Gasperini.

  Papini, Wood and Cai [24] showed
that  the same maximal acceleration (10) of an extended particle
follows naturally from the theory of conformal transformations. On
the other side, they  also studied [25] the motion, in a sort of
Madelung fluid, of a spherically symmetric extended object, a bubble,
of Riemannian geometry embedded in external Weyl geometry where a
conformal covariant calculus is used. The field equations for that
case are obtained starting from the conformally invariant action $$
I_{C} = \int \{ - \frac{1}{4} f_{\mu \nu} f^{\mu \nu} + |\beta|^{2}
\bar{R} + k |\barbox_{\mu} \beta \barbox^{\mu} \beta| + \lambda
|\beta|^{4} \} \sqrt{-g} d^{4} x $$ $$+  \int \rho g^{\mu \nu}
\gamma_{\mu} (\barbox_{\nu} \rho - \epsilon \rho \varphi_{, \nu} )
\sqrt{- g} d^{4} x,  \eqno(11)$$ where $\beta$ is a complex scalar
field with $ \rho = |\beta|$ and $\varphi = arg \beta$, then
$\gamma_{\mu}$ is a vector Lagrange multiplier and $k$ and $\lambda$
are arbitrary constants. Furthermore an overbar is used to
distinguish an object defined in terms of the gauge-covariant
calculus of Weyl geometry from the corresponding object associated
with the covariant calculus of Riemannian geometry. For example,
$\stackrel{-}{\Box}$ is the spacetime gauge-covariant derivative.
 Even in this very different case they found a maximal
acceleration for the bubble  $$ A_{max} =\frac{2}{R} \eqno(12)$$
(where R is the bubble's radius) very similar to the value (10).

{\it In any case we can conclude that the maximal acceleration
depends on the characteristic size of the extended object we are
considering}. \vspace{.3truecm}

 \section{Maximal acceleration for classical charged particles}
Recently Goto [26] has studied the equation of motion for a classical
charged particle including radiation reaction force performing a
stable hyperbolic motion immersed in a uniform gravitational field.
He finds that the observer in the laboratory frame measures the
gravitational force acting on charged particles as $$ F_g =
\frac{mg}{1-g^2 \tau ^{2}} \eqno(13)$$ where $$ \tau =
\frac{2e^2}{3mc^4} \eqno(14)$$ His interpretation of equation (13) is
 that the gravitational mass of charged particles should be slightly
greater than its inertial mass. But we can write his formula also as
$$ F_g = \frac{mg}{1-g^2/A^{2}_{max}} \eqno(15)$$ and interpret it
considering that an infinite gravitational force $F_g$ is necessary
to produce an acceleration $$g = A_{max} = \frac{3mc^4}{2e^2}
\eqno(16)$$ {\it In this case the maximal acceleration depends on the
charge of the classical particle}.

It is worth noting that Caldirola [27], in his theory of the
classical electron founded on the introduction of a fundamental
interval of time (the so called chronon), showed the existence of a
maximal value of acceleration equal to half of  Goto's one (16).

\section{Maximal acceleration for nonspreading wave packets}

Recently Caldas and Silva [28] have analyzed the motion of a
nonspreading
 wave packet in a harmonic potential. It is well known that in the
 case of a quantum harmonic oscillator the motion of the center of a
 wave packet is rigorously identical to that of a classical particle.
 Caldas and Silva impose that the packet does not spread
 so that $$ [\Delta q(t)]^2_{time average} = [\Delta q_o]^2 \eqno(17)$$ from
 which they obtain [28] $$ \Delta p_o = m \omega \Delta q_o \eqno(18)$$
and assume that, at the initial time, the wave packet is such that $$
\Delta q_o \Delta p_o = \frac{\hbar}{2} \eqno(19)$$ This way $$m
\omega (\Delta q_o)^2 = \hbar /2 \eqno(20)$$ But considering that the
classical particle (hence the peak of wave packet) obeys $q(t) = A
cos\omega t $,  we have $$v_{max} = \omega A \leq c \eqno(21)$$ and
$$ a_{max} = \omega^2 A \eqno(22) $$ From (20) (21) and (22) we
obtain the relation: $$a_{max} \leq {\hbar c \over 2m \ell^2 }
\eqno(23)$$ where $\ell$ is the characteristic size of the packet
$\Delta q_o$. {\it In this case we find a maximal acceleration that
depends on the square length of the wave packet}.

 On the contrary,
Caldas and Silva calculate a "classical variance" $ \Delta q = A^2
/2$ and require that it can be identified with the quantum variance,
obtaining from (20) $$ m\omega A^2 = \hbar\eqno(24)$$ From (22) and
(24) and putting $\omega A = c $ they find a maximal driving force
that we can read as a maximal acceleration $$A_{max} = {m c^3 \over
\hbar} \eqno(25)$$ similar to Caianiello's one (3).
\section{Maximal acceleration from maximal temperature}

It is very easy to derive the existence of a maximal acceleration
from a maximal temperature by using Unruh and Davies demonstration
[29] stating that a particle-detector
 subject to a constant acceleration would react to vacuum fluctuations
  as if it were at rest within a gas of particles having a temperature
   proportional to acceleration
    $$T = {{\hbar a}\over{2 \pi k c}} \eqno(26)$$
where $k$ is Boltzmann constant.

 Brandt [30], for example,  starts from the result obtained by Sakharov
 [31],
  according to which the absolute temperature of thermal radiation
in vacuum is $$T_{max} \simeq {{c^2}\over{k}} \sqrt{{\hbar
c}\over{G}} \eqno(27)$$ that in (26) implies that there is a maximal
acceleration relative to vacuum: $$ A_{max} \simeq
\sqrt{{c^7}\over{\hbar G}} = {{m_p c^3}\over{\hbar}} \eqno(28)$$  It
is worth noting that $ A_{max}$ is similar to maximal acceleration
found by Caianiello, but in this case the rest mass is substituted by
the Planck mass $m_p = (\hbar c/G)^{1/2}$ {\it In this case the
maximal acceleration is a universal constant} and assumes an
extraordinarily high value: $A_{max} \simeq 5 \times 10^{53} cm/s^2
$. Therefore it is very difficult to find through experimental tests
some physical effects which can be ascribed to the existence of this
upper limit on the acceleration.

The same demonstration can be performed starting with other values of
maximal temperature available in literature [32]. For example,
another interesting critical value of the temperature is the so
called Hagedorn temperature $T_H \propto \alpha^{\prime -1/2} $ that
arises also in string thermodynamics. Parentani and Potting [33]
studied the motion of a string in Rindler frame and found the
occurrence of a maximal temperature $T_{max} = T_H/\pi$ above which
the string partition function diverges. Substituting this value of
maximal temperature in the Unruh formula (26), we can find the same
maximal acceleration of Gasperini.

 \section{Conclusions}
  The maximal acceleration principle can be successfully used to prevent
the occurrence of singularities in General Relativity [16 - 19], and
of ultraviolet divergences in quantum field theory [34], in
particular in the estimation of free energy and entropy of quantum
fields [35]. We have shown that several possible values of maximal
acceleration can be found. The choice among them is crucial to obtain
the right model of a relativistic dynamics with an upper limit on the
acceleration and can be definitely done only through experiments.
Finally it is even possible that two different values of maximal
acceleration can survive in the same model. Using in (1) Planck
acceleration (28), we obtained [36] a modified Rindler metric $$
ds^{2}  = -(\xi^{2} - A^{-2}_m)    d \eta^{2}    +   d   \xi^{2} + d
y^2 + d z^{2}\eqno(29)$$ We showed that, in the case analyzed by
Frolov and Sanchez, a maximal acceleration depending on the string's
length $\lambda$  still exists and it does not diverge in the limit
$\lambda \rightarrow 0$, but we have $a \rightarrow A_m = m_p
c^3/\hbar$. {\it As in classical relativity only particles with zero
mass can move at the maximal velocity $c$, so in our theory only
point particles can move at maximal acceleration $A_m$.}

\section*{Acknowledgement}
The author is grateful to Gaetano Scarpetta and Giorgio Papini for
useful suggestions.

\section*{References}
\begin{enumerate}

 \item  H. Weyl, {\it
 Sitzungsberichte der Preussinske Akademie der Wissenschaften},
{\bf 465}, (1918)

\item G. Amelino-Camelia, {\it Int. J. Mod. Phys.} {D 11}, 35 (2002);
{\it Phys. Lett.} {\bf B 510}, 255 (2001)

\item J.Magueijo and L.Smolin, {\it Phys. Rev. Lett.} {\bf 88},
190403 (2002)

\item S.G. Low,  {\it J. Math. Phys.}, {\bf 38}, 2197 (1997)

\item D.V.Ahluwalia, {\it Phys. Lett}, {\bf A 275}, 31 (2000);
D.V.Ahluwalia and M.Kirchbach, "Fermions, bosons and locality in
special relativity with two invariant scales" in gr-qc/0207004

\item A.A. Ketsaris, "Accelerated motion and special relativity
transformations" in physics/9907037 and "Turns and special relativity
transformations" in physics/9909026

\item G. Amelino-Camelia, "Status of Relativity with observer
independent length and velocity scales" in gr-qc/0106004

\item E.R.Caianiello, {\it Nuovo Cimento}, {\bf 59B}, 350 (1980);
 E.R.Caianiello, G.Marmo, G.Scarpetta, {\it Nuovo Cimento}, {\bf 86A}, 337
(1985)

\item E.R.Caianiello, {\it Lett. Nuovo Cimento}, {\bf 32}, 65 (1981)

\item G.Scarpetta, {\it Lett. Nuovo Cimento}, {\bf 41}, 51 (1984); W.Guz,
 G.Scarpetta, in {\it  Quantum field theory}   ed. F.Mancini
      (North-Holland, Amsterdam,1986), p.233

\item H.E. Brandt, {\it Found. Phys. Lett.}, {\bf 2}, 39, (1989)

\item F.P. Schuller, {\it Annals Phys.}, {\bf 299}, 174, (2002)

\item E.R. Caianiello, A.Feoli, M.Gasperini and G.Scarpetta,
 {\it International Journal of Theoretical Physics}, {\bf 29}, 131 (1990)

\item  E.R. Caianiello, {\it Lett. Nuovo Cimento}, {\bf 41}, 370,
(1984); W.R. Wood, G. Papini, and Y.Q. Cai, {\it Nuovo Cimento}, {\bf
104 B}, 361, (1989)

\item
 A.Feoli, G. Lambiase, G.Papini and G.Scarpetta,
{\it Phys. Lett.}  {\bf A263}, 147 (1999); S.Capozziello, A.Feoli,
G.Lambiase, G.Papini and G.Scarpetta, {\it Phys. Lett.} {\bf A268},
247 (2000).

\item
 V.Bozza, A.Feoli, G.Papini and G.Scarpetta,
{\it Phys. Lett.} {\bf A271}  35 (2000).

\item
V.Bozza, A.Feoli, G.Lambiase, G.Papini and G.Scarpetta, {\it Phys.
Lett.}  {\bf A283}  53  (2001)

\item E.R. Caianiello, M. Gasperini and G. Scarpetta,
 {\it Class. Quan. Grav.} {\bf 8}, 659 (1991);
 E.R.Caianiello, A.Feoli, M.Gasperini and G.Scarpetta,
 ``Cosmological implications of maximal proper acceleration''
 in ``Bogoliubovskie \v{c}tenija'', eds. E.V.Ivaskevic
 and E.V.Kalinnikova, (Joint
Institute for Nuclear Research, Dubna, 1993) p.134

\item
V.V. Nesterenko, A.Feoli and G.Scarpetta, {\it Jour. Math. Phys.}
{\bf 36}, 5552 (1995);
 {\it Class. Quantum Grav.} {\bf
13}, 1201 (1996)

\item C.W.Misner, K.S.Thorne and J.A.Wheeler, {\it Gravitation} (W.H. Freeman
and Company, San Francisco, 1973), chapt. 6

\item N.Sanchez and G.Veneziano, {\it Nucl. Phys.}, {\bf B333}, 253 (1990);
M.Gasperini, N.Sanchez and G.Veneziano, {\it Nucl. Phys.}, {\bf
B364}, 365 (1991); {\it Int. J. Mod. Phys.}, {\bf A6}, 3853 (1991)

\item M.Gasperini, {\it Phys. Lett.}, {\bf B258}, 70 (1991);
 {\it Gen. Rel. Grav.}, {\bf 24}, 219 (1992)

\item V.P.Frolov and N.Sanchez, {\it Nucl. Phys.}, {\bf B349}, 815 (1991)

\item  R.W.Wood, G.Papini and Y.Q.Cai, {\it Nuovo Cimento}, {\bf B104}, 653 (1989)

\item G. Papini and W.R. Wood, {\it Phys. Lett.} {\bf A170},
409(1992).

\item M. Goto, "The equivalence principle and
 gravitational and inertial mass relation of
classical charged particle" in physics/0104021

\item P.Caldirola, {\it Lett. Nuovo Cimento} {\bf 32}, 264 (1981)

\item H.C.G. Caldas and P.R. Silva, {\it Apeiron} {\bf 8} n.1 (2001)
(gr-qc/9809007)

\item P.C.W.  Davies,     {\it J.  Phys.   A.}, {\bf 8},  609 (1975);
 W.G.    Unruh, {\it Phys.   Rev.}, {\bf  D14},    870 (1976).

\item  H.E. Brandt, {\it Lett. Nuovo Cimento}, {\bf 38}, 522, (1983)

\item A.D.    Sakharov,
    {\it JEPT    Lett.}, {\bf 3},  288 (1966).

\item K.Huang and S.Weinberg, {\it Phys. Rev. Lett.} {\bf 25}, 895 (1970)

\item R. Parentani and R. Potting {\it Phys. Rev. Lett.} {\bf 63}, 945 (1989)

\item
V.V.Nesterenko, A.Feoli, G. Lambiase and G.Scarpetta,
 {\it Phys. Rev.} {\bf D 60}, 65001 (1999).

\item M.McGuigan, {\it Phys. Rev.} {\bf D 50}, 5225, (1994)

\item
A.Feoli, {\it Nuclear Physics}  {\bf B396}, 261 (1993)

\end{enumerate}

\end{document}